\documentclass[sn-basic]{sn-jnl}

\jyear{2023}%

\theoremstyle{thmstyleone}%
\theoremstyle{thmstyletwo}%

\theoremstyle{thmstylethree}%

\begin{document}

\title[Review Bomb]{Ideology-driven polarisation in online ratings: the review bombing of The Last of Us Part II}


\author*[1]{\fnm{Giulio Giacomo} \sur{Cantone}}\email{prgcan@gmail.com}

\author[2]{\fnm{Venera} \sur{Tomaselli}}

\author{\fnm{Valeria} \sur{Mazzeo}}

\affil[1]{\orgdiv{Science Policy Research Unit}, \orgname{University of Sussex}}

\affil[2]{\orgdiv{Department of Economics and Business}, \orgname{University of Catania}}


\abstract{
A review bomb is a large and quick surge in online reviews about a product, service, or business, coordinated by a group of people willing to manipulate public opinion about that entity. This study challenges the assumption that review bombing is solely a phenomenon of misinformation and connects motivations and substantial content of online reviews with the broader theory of judgement of facts and of value. These theories are verified in a quantitative analysis of the most prominent case of review bombing, which involves the video game The Last of Us Part II. It is discovered that ideology-driven ratings are followed by a grassroots counter-bombing, aimed at mitigating the effects of the negative ratings. The two factions of bombers, despite being politically polar opposites, are very similar in terms of other metrics. Evidence suggests the theoretical framework of political disinformation is insufficient to explain this case of review bombing. In light of the need to re-frame review bombing, recommendations are proposed for the preventive management of future cases. }



\maketitle

\section{Introduction}

Platforms of online reviews are websites that allow people to sign up \textit{accounts} and contribute to the collective production of the \textit{content} of the website itself. Users can \textit{influence}, supposedly for good, the behaviour of others, providing useful information through submitting new content in the platform. In other words, people share personal experiences and opinions on specific products, services, or topics, and in doing so influence the decision-making process of other users. These platforms have become one of the main sources consumers rely on to retrieve information needed for deciding about purchases \citep{watson_swayed_2018,stockli_recommendation_2021,watson_impact_2022,sharkey_expert_2023}.


Platforms often implement a system of online reviews as a core function of their own business idea (e.g. Tripadvisor) or as a system to support the consumer's decisions (e.g., Amazon, Netflix), the latter being referred to as a form of recommender system. People will register a \textit{user} account and will be provided with the right to rate \textit{items} in a catalogue. These items may be movies, products, services, or even people or ideas. The overall process of explicit evaluation, or `rating', happens typically through a combination of two methods: (i) a numeric score or (ii) a textual comment, which is the `review'. In recommender systems data are often integrated with other implicit evaluations, i.e. metrics on user behaviour like counts of particular actions \citep{aggarwal_recommender_2016,stockli_recommendation_2021}.

There is an extensive debate on the effectiveness of online ratings to infer real traits of the \textit{items} they target. There is a fundamental problem in assuming that a sample of scores or texts is representative of general opinions on the traits of the items: differently from experimental trials, platforms have almost no control over who will rate the items. The problem is particularly well-understood in a comparison of two or more items or in pooling ratings on the same item across many platforms: the samples of ratings come from data-generating processes that have substantial differences. For example, a less-known item may receive mostly ratings from people with a taste for a niche (having a ``hipster" taste); the same people will be less condescending in rating a more general item. To make it simple: the comparison of two samples of ratings (or reviews) could end in `comparing apples with oranges' because the populations of raters differ. Conclusive empirical evidence shows that online ratings display a peculiar J-shape, instead of a canonical bell curve, which results in experimental trials instead \citep{li_self-selection_2008,hu_overcoming_2009,han_customer_2020,smironva_self-selection_2020,brandes_extremity_2022}. Under these premises, it is not obvious how well canonical synthetic measures, i.e. average scores or results from sentiment analysis, are representative of true traits of the items.

Online reviews have other drawbacks: reviews can be faked, and scores can be unfairly biased. Under which circumstances would a consistent portion of evaluations on an item be inauthentic? To provide an example, a business could try to discredit competitors, but these can always retaliate. In the end, gossip would propagate inauthentic ratings within the platform. This is more a theoretical model than a factual representation of what research observed. More commonly, businesses compete only by boosting their own reputation with `astroturf' campaigns, where `sock-puppets', which are inauthentic user accounts (e.g. a `bot network' of fake accounts) following the order of a master, promote the product posing as real consumers of the product \citep{ratkiewicz_truthy_2011,anderson_reviews_2014,mayzlin_promotional_2014,lappas_impact_2016,luca_fake_2016,kumar_army_2017,zhuang_manufactured_2018,wu_fake_2020,zhao_cross-site_2021,petrescu_innocent_2022,petrescu_man_2023,wang_benefits_2023}.

\subsection{Types of judgements and Review Bombing}

These problems are structurally linked to the methods of accessing the role of `rater' of the items. Who should be entitled to be a rater? A verified account? Someone who has a lot of experience? Or whoever is good-willed enough to provide their own opinion? As in many social media, there is clearly a trade-off between the scale of the data-generating process (in collecting ratings), and the trustfulness of its summary information. Platforms with restricted access (e.g., to those who pay a fee for use) are not as much biased as `free to use' platforms. However, are the latter those of particular interest: open platforms can have populations of raters more representative of the general diversity of opinions in the public, and this diversity can definitely prove its usefulness. While the management of these information systems pertains to the sphere of the private, online reviews act also as public goods: they are both a source of information for everybody and digital spaces where opinions can be expressed, too.

Referring back to the second point, it is essential to emphasize that, in recognizing online reviews as a means of public space for people's judgement, it has been overlooked a significant distinction historically addressed by many classical authors such as David Hume, Immanuel Kant, Max Weber, and Herbert Simon. This distinction pertains to the difference between judgements of fact, or technical judgements, and judgements of value, or moral judgements \citep{fuchs_observing_2017,stewart_evaluative_2017}. An example: one can use an online review to voice criticism against a firm perceived as unethical. It is not a technical (fact-based) evaluation, but a moralistic (value-based) one. Fact-based judgements are rooted in the desirability to reach objectivity and unbiased comparability between natural objects or artefacts. These proprieties spouse well with the task of synthesising multiple opinions through quantitative procedures. Value-based judgements are rooted in the desirability of the outcomes of human choices. They have to do with conformity to general or local morality, and more in detail with ideological politics and group identities, too. 

The concept of Review Bombing (RB) is pivotal for understanding the epistemological problem that value-based reviews pose in the information management of online reviews. In an RB there is a whole `mob' of accounts (figuratively, a digital crowd) that leave a large number of negative reviews and low rating scores in a short time span. Compared to a dislocated collection of value-based judgements, a collective RB is more effective in signalling the existence of moral distress towards the item. The immediate goals of a review bomber are to artificially lower the product's average rating and harm the reputation of the item, but bombers are often motivated by the belief that their collective action can stimulate the interest of the media and finally discourage the consumption of their target.

There is a connection between RB and fake reviews: since they operate in open, unrestricted, digital spaces, bombers can definitely submit a lot of inauthentic evaluations or even resort to \textit{sock-puppets} to reach their goal of flooding the item's web page with negative ratings. In other words, RB is a phenomenology where a moralistic criticism of an item can coexist with unfair (and, possibly, \textit{unethical}!) behaviour towards it. Concerns from the management of online platforms are deserved, especially when bombers intend it as a form of cultural jamming aimed at subverting the use of the platform\footnote{Invited to a lecture in New York in 1967, Umberto Eco proposed the expression `semiological guerrilla', which suggests an imaginary of cheap-equipped people recurring to dirty tactics to subvert a cultural establishment \citep{eco_travels_1990}.}: such actions undermine the platform's credibility as an informative tool.

\subsection{Aims and findings}

Commonly proposed solutions to the conflict between the platform and the bombers often are limited to plan when to suppress the RB through the removal or the shrinkage of value-based ratings in the processes of inference and ranking \citep{luca_fake_2016,li_bimodal_2017,volkova_misleading_2018}. There is, therefore, a temptation to conflate the bias from ideologically-charged reviews to the bias induced by fake content and spam reviews. These solutions are coherent with the paradigm of online platforms as inferential tools but negate their role as digital spaces of debate.

Is suppression an overall effective strategy? Hardly this question has an affirmative answer. Once the RB is acknowledged by the media, the reputational damage to the platform is already done but a premature intervention may be perceived as unjustified interference and as a sign of confirmation of the lack of ideological neutrality of the evaluative tool towards the object of evaluation \citep{graves_anatomy_2017}.

In general, there is a gap in the literature about non-invasive solutions to ease the latent conflict between the platform management and the bombers, because RB has not been studied extensively. One of the aims of this study is to identify managerial recommendations that do not rely exclusively on the suppression of information, but are geared towards preventing excessive distortion of the scores and conveying the pre-existing tension of bombers in a more useful way while demonstrating the platform's commitment to transparency and fairness.

In order to achieve this primary goal, this study proposes to not solely focus on automatic detection methods for deception but rather to understand the root causes and global effects of RB. One might be tempted to assume that since people rely on online reviews, bombers genuinely believe that their actions can change public opinion. However, it is possible that bombers are driven by moral and ideological motivations, and do not have any expectation of being socially influential, but rather boycott a product out of an emotional outburst or because they believe it is the right thing to do \citep{eddy_perception_2020,altay_people_2023,altay_misinformation_2023}. These hypotheses are typically overlooked in the literature on online deception but are important distinctions for understanding how platforms can prevent Review Bombing, rather than simply suppressing it \textit{ex post facto}.

To investigate these secondary goals, the section Materials provides a quantitative overview of the Review Bomb of the video game ``The Last of Us Part II" (TLOU2) on the website Metacritic, possibly the most extended and relevant case of RB. Public reviews within the first weeks after the release are analysed through three metrics: distribution of the scores, the lexical richness of the reviews, and the number of previous reviews on Metacritic by reviewers of TLOU2. The major result of this analysis is that RB has been followed by an ideologically polar opposite movement of `anti-bombers'. These are ideologically against the bombers because they label them explicitly as a source of misinformation and bigotry. However, $\sim$ 80\% of all the users involved in reviewing TLOU2 never reviewed any other item before it, nor did they review other items in the months following their review of TLOU2, with no meaningful differences between the two factions. Both factions, while keeping antithetical opinions, behave similarly in terms of lexical richness, too. Some reviews leave the suspicion of being at least partially inauthentic, but the overall behaviour of these users does not predominantly fit cues associated with \textit{sock-puppets}.

The discovery of polarisation in reviews leads naturally to further semantic analysis in the section Thematic Labelling: prevalence and sentiment of ideological themes driving the Review Bomb are quantified through a systematic procedure that involves the identification of relevant semantic keywords within the reviews. This procedure led to the further discovery of a new category of theoretical interest, labelled `meta-opinion'. It refers to reviews motivated by the goal to support or to rebut another opinion, even in absence of any judgement on the object of the debate (in this case, the technical merits of the video game). Meta-opinions seem the main driver of the growth of anti-bombers, and a meaningful discovery is that this effect is larger than what can be attributed to homophobia or political grudge, for bombers.

Noticing the prevalence of meta-opinions has provided valuable insights for formulating recommendations on how to enhance the functionalities of online review platforms, which are collected in the section Discussion. The Conclusions briefly discuss some limitations of this observational study and propose future research directions. In particular, the distinction between fact-based and value-based judgements is revisited, and it is suggested that online reviews should receive greater attention from scholars in the theory of evaluation, as they may be the most fertile ground for the development of mixed evaluative methods.

\section{Materials}

\textit{Metacritic} is a website that collects professional reviews for movies, TV shows, music, and video games. They developed the METASCORE algorithm, which aggregates the opinion of experts into a unique number within $[0:100]$. Metacritic also allows registered users to assign a rating score within $[0:10]$, and the sample average of the scores is called \textit{User Score} \citep{drachen_only_2011,johnson_edge_2014,kasper_role_2019,santos_whats_2019}.

\begin{figure}[hbt!]
	\centering
		\includegraphics[width=11cm, height=6cm]{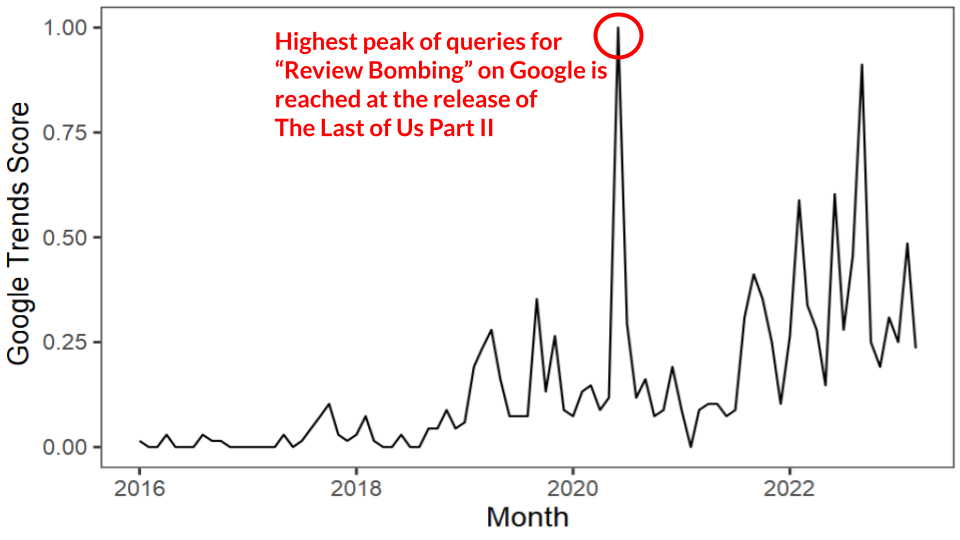}
	\caption{Search trends in Google Trend for \lq review bombing' on Google. It spikes in searches in June 2020, the release date of The Last Of Us Part II.}
	\label{fig:bomb}
\end{figure}

On June 19$^{th}$, 2020, Sony Entertainment released the horror-themed video game \textit{The Last of Us Part II} (TLOU2). Just a few days after, TLOU2 became the most-reviewed item on Metacritic. During the first days, the User Score floated towards values below 5/10. After 40 days it stabilised around a value of 5.7, after having reached $\sim$ 65k reviews. A User Score of 5.7 is considerably low for Metacritic (7 can be assumed as an average User Score). This is the widest case of \lq Review Bombing' (RB) until then (see, Fig. \ref{fig:bomb}).

According to journalistic coverage, the RB of TLOU2 has been pre-emptively coordinated by a group of ideologically-driven users. Reported causes of the grudge among these users are:

\begin{itemize}
\item Disappointment with artistic choices regarding the artistic direction for the protagonist of the prequel ("The Last of Us").
The decision to feature two female protagonists, one of whom is a female bodybuilder, has been met with disappointment by those who view this artistic choice as a forced attempt to fit a feminist narrative within the horror genre. Additionally, the inclusion of LGBTQ (Lesbian, Gay, Bisexual, Transsexual, Queer) romance has been perceived as challenging the established cultural identity of the typical player of horror-themed video games.
Reluctance to accept LGBTQ main characters in fiction for entertainment is often associated with conservative and far-right politics in the United States, where some political groups fear that the entertainment industry spouses a progressive cultural agenda, influencing younger generations \citep{dutton_digital_2011,filip-crawford_homosexuality_2016,young_media_2017}.

Part of these choices has been already disclosed by trailers of the video game or similar sources of advertising campaigns in the pre-release. This gave time for bombers to organise themselves, likely on the website 4chan or similar anonymous online venues.
\item Video game players who identify with far-right ideologies are also associated with supporting \#GamerGate \citep{braithwaite_its_2016}, an opinion movement born on the website 4chan that alleges biases of many natures in the journalistic coverage of video games \citep{ferguson_who_2021}. These supporters of \#GamerGate, being stimulated by the pre-release teasers, had the time to fabricate a shared narrative around TLOU2 being an example of intellectual dishonesty and moral abjection in the video game industry.
\end{itemize}

The debate surrounding TLOU2's alleged over-politicisation is part of a larger historical trend of criticism towards the artistic direction of media entertainment. The debate holds some level of legitimacy in more conservative circles. Sometimes it veers towards conspiratorial thinking, particularly when extreme statements are claimed without supporting evidence. Remarkable is the continuity of such theories to arguments that are originally proposed in an online forum like 4chan, which has been associated with the spread of misinformation and \lq alternative facts" including through the controversial political campaign of former President of United States Donald Trump \citep{hine_kek_2017,tuters_post-truth_2018,ludemann_digital_2021,zelenkauskaite_shades_2021}.

On Metacritic the submission of a score only (no textual review) happens anonymously: the information about the individual score of the user to the item cannot be retrieved. However, if a user also provides a textual review (up to 5,000 characters), then the review, the score, the day of submission, and the username can be retrieved from the dedicated webpage for the item on Metacritic. In the next sub-section are summarised informative metrics about the first three weeks of public reviews on TLOU2.

\subsection{Summary}

Data collection has been conducted in two rounds. In the first round, operated at August $1^st$ 2020, 65,466 reviews have been scraped with software \texttt{Rvest} \citep{wickham_rvest_2022} from the public Metacritic web page for \textit{The Last of Us Part II}. For each public review, the username of the author ($u$), a textual comment on TLOU2, a score ($x$), and a date were collected.

More than 20 different languages have been detected using software for language detection  (One round of \texttt{langdetect} in Python, then one round of \texttt{cld3} in R). Manual checks have been performed for non-matching results. 78\% of the reviews were written in English. Many languages had only a trivial frequency in the sample. In Table \ref{tab:lang} languages with less than 250 reviews are conveniently grouped according to geography and culture. Some reviews consisted only of symbols (for example, emoticons) and were grouped into ``Others", too.

\begin{table}[hbt!]
\caption{Statistics on languages.}
\label{tab:lang} 
\centering
\begin{tabular}{p{6.5cm}p{1.6cm}p{1.0cm}}
\hline\noalign{\smallskip}
Language & n & $\bar{x}$\\
\noalign{\smallskip}\hline\noalign{\smallskip}
English & 51,120 & 5.0 \\
Spanish & 5,882 & 6.1 \\
Russian, Greek and East European & 3,419 & 6.0 \\
Portuguese & 3,408 & 7.3 \\
Chinese, Japanese, Other Asians & 351 & 3.0 \\
Italian & 340 & 8.3 \\
German, Baltics and Nordics & 284 & 8.0 \\
French & 275 & 7.6 \\
Arabic, Farsi and Turkish & 260 & 5.3 \\
Others and only symbols & 127 & 4.7  \\
\noalign{\smallskip}\hline\noalign{\smallskip}
\end{tabular}
\end{table}

Table \ref{tab:lang} provides non-conclusive evidence for the hypothesis that the artistic direction TLOU2 has been received differently across different cultures. The over-representation of reviews in English suggests that non-English reviews may not be representative of the Metacritic user population in the associated geographic regions. For example, those who chose to write in their native language may have had more nationalistic tendencies than those who opted to write in English, and this influenced their perception of the controversy regarding the video game. It is not assumed that users who wrote in English were all native English, too. Furthermore, differences in the mean score ($\bar{x}$) across languages cannot be solely attributed to cultural predispositions towards TLOU2, as there is no counterfactual information about general rating tendencies for video games.

To ensure a cohesive textual representation of the RB on TLOU2, all subsequent descriptions and analyses are conducted exclusively on the 51,120 reviews in the English corpus.

\subsubsection{Experienced users}
The second round happened 6 months later, at $16^t$$^h$ January 2021: the number of past reviews on Metacritic ($k$) has been scraped from each account of those users who authored a review in English. Only 21\% ($n = 10,552$) had already submitted at least another review ($K = 1$). The other 79\% ($n = 40,568$) are accounts registered exclusively to review TLOU2 ($K = 0$)\footnote{$k$ has been scraped immediately after the first round of scraping. Only $1,480$ users of the $K=0$ group made at least a review after TLOU2, so this intermediate round will be ignored, and the $k$ recorded in January is adopted.}. It is assumed that users in the $K=1$ group are more experienced than the majority of the `causal' reviewers ($K=0$), meaning that they have \textit{previous experience} with Metacritic.

Average scores differ between the first group ($\bar{x}_{K=1} = 4.63$) and the second ($\bar{x}_{K=0} = 5.15$). This difference, although small, goes in the direction to think that review bombers, who tactically assign $x < 2$ to lower the User Score, did not abuse of \textit{sock-puppets}; at least not more than `not-review-bombers'.

\subsection{Detection of fakes and lexical diversity}

In the English corpus, there are at least 1,722 very suspicious reviews. These are detected through cues:

\begin{enumerate}
\item username made only of numbers (e.g., ``221000000").
\item high-similarity among two or more usernames (e.g., ``david2000" and ``davvid2000") or high-similarity in the text\footnote{String matching (through Python's library \texttt{fuzzywuzzy}) has been applied to determine the level of similarity between texts and usernames. The matching algorithm has been based on the minimisation of Levenshtein distance, that is the minimum number of atomic edits (additions, deletions, or substitutions) required to change one string into another \citep{chen_email_2014}.}.
\item the same token is repeated more than 2 times in a row in a review (e.g., ``this game is horrible horrible horrible").
\item the same character is repeated more than 3 times in a row in a review (e.g., ``aaaaaah this game is horrible")
\end{enumerate}

The assumption behind cue 2. is that fake reviews can be identified because their content is fake, or because the user account is fake, and that in particular specific expressions and lexical constructs are recurrent \citep{mewada_research_2022}. However, the primary assumption beyond all cues is that since review on Metacritic has a minimal requirement to be made of at least 75 characters, then abusers may resort to copying and pasting their own text or to write nonsense just to meet this requirement quickly and leave multiple reviews using \textit{sock-puppets} accounts.

The finding that only 1,722 accounts (less than 4\% of the total) may be associated with fake reviews is lower than expected. It must be remarked that well-fabricated fake reviews can be very difficult to distinguish from authentic ones. Another hypothesis is that Metacritic already suppressed reviews before August. Therefore, this result alone should not be taken as conclusive evidence. A more detailed analysis of fake reviews is presented in Appendix D.

\subsubsection{Lexical diversity}

Generally, suspicious texts can have a high number of characters (\textit{nchar}) but lack any meaningful content. For example, the text could just be a long repetition of an ideological (often rude) slogan. To address this issue in measuring the length of the text, the lexical diversity ($D$) metric \citep{zhou_survey_2020} is employed instead. $D$ and \texttt{nchar} are highly correlated (Pearson's correlation $=0.96$).

\subsection{Polarisation of the rating scores}

\begin{figure}[hbt!]
	\centering
		\includegraphics[width=12cm] {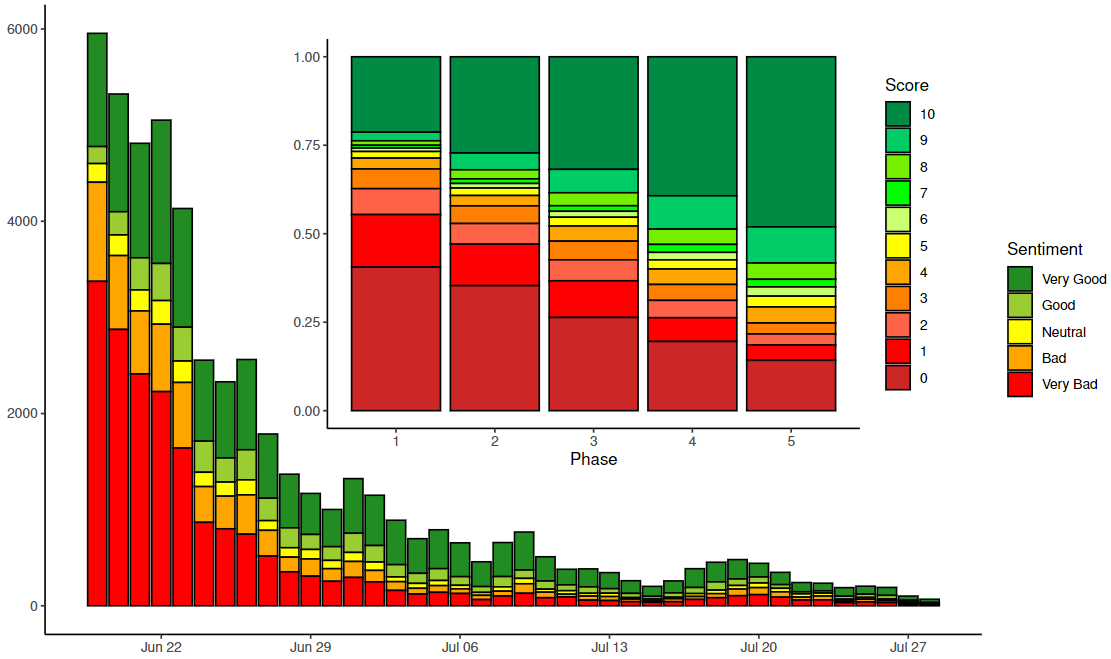}
	\caption{\textbf{Comparison of rating sentiments trending in absolute values in the 40 days.} \\ The label Sentiment aggregates rating scores (x) according the following rules: \\ Very Bad = [0:1]; Bad = [2:4]; Neutral = [5:7]; Good = [8:9]; Very Good = 10. \\ In the inner figure are represented relative frequencies of rating scores in five consecutive phases of an equal number of reviews ($\sim10,000$ reviews per phase) but unequal time duration. \\ Phase 1 consists in only the first 2 days (19th and 20th June), and Phase 2 in the next two (21th and 22th June). Phase 3 lasts from 23th June to 27th June. Phase 4 lasts from 28th June to July 2nd. From July 3rd to July 28th is the last phase. \\ This aggregation shows the relative decline over time of low scores (red color) and the rise of $x = 9$ as a score, alongside $x = 10$, which is already prevalent among higher scores (green color) in the first phases.}
	\label{fig:time}
\end{figure}

Fig. \ref{fig:time} shows that $\sim 70\%$ of the scores ($x$) were of $0$, $1$, or $10$. A bi-polar distribution of frequencies is typical of online review scores but compared to statistics provided by Schoenmueller et al. \citep{schoenmueller_polarity_2020}, the frequency distribution of TLOU2's scores in Table \ref{tab:scores} displays an inflated bi-polarity.

The polarisation of the scores happened asynchronously (Fig. \ref{fig:time}): an early wave of very negative reviews, concentrated in the first $4\sim6$ days, then a wave of very positive reviews spread more equally over time, and prevalent in the latter two phases. The concentration on the first days is a typical pattern observed in online reviews: people try the product soon after its release and feel more motivated to write a review to participate in the ongoing discussion about it.

Reading a few of these very positive reviews it can be noticed that many reviewers felt the motivation to rate TLOU2 $x = 10$ to balance out the unfair User Score resulting from the RB. In some cases, they explicitly mention that they do not think that, on a technical level, TLOU2 deserves a `10/10', so they \textit{consciously} added a positive bias because they feel morally aligned with the video game, or against the negative bombers.

\begin{figure}[hbt!]
	\centering
		\includegraphics[width=12cm] {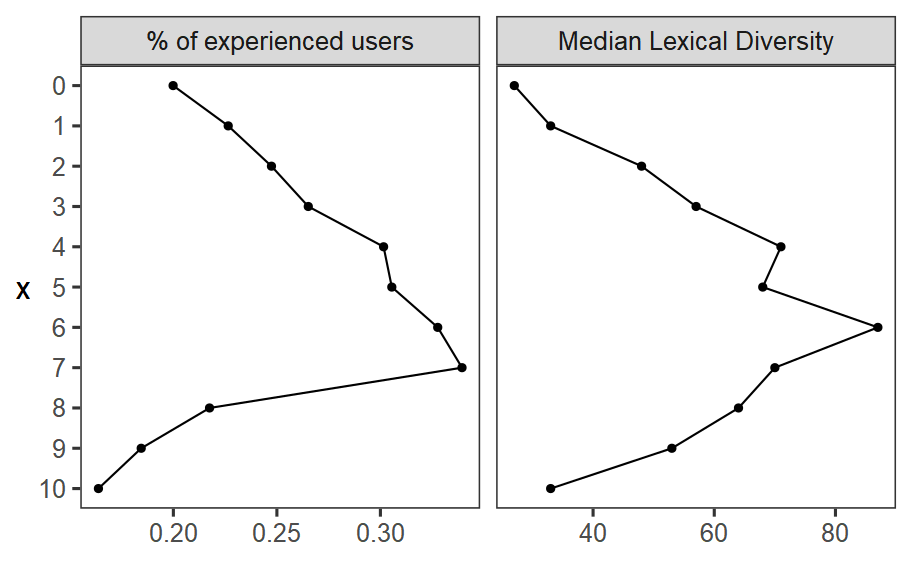}
	\caption{\textbf{Crossing scores with \% of $K=1$ and with Median $D$}}
	\label{fig:conc}
\end{figure}

The analysis of the distribution of lexical diversity and the relative frequency of $K=1$ in Fig \ref{fig:conc} reveals two concave curves, with the highest values observed at central scores, and lower values at the extreme scores $x$. Indeed, $x = 10$ displays a lexical diversity level similar to that of $x < 2$.

The implication is that while arguments of bombers and anti-bombers were polar opposites on the ideological ground, their engagement and their expertise as reviewers \citep{santos_whats_2019} were similar. In both cases, this $\sim 70\%$ of users who adopted a scoring behaviour coherent with the hypothesis of being motivated to tactically alter the User Scores on average look less qualified or less engaged in Metacritic than the remaining $\sim 30\%$ of users. Numerical details of Fig. \ref{fig:time} and Fig. \ref{fig:conc} are provided in Appendix A.

\section{Thematic Labelling}

In Materials has been established that a wave of very positive reviews followed a RB, and it has been regarded as an anti-RB. It has been discovered that on experience with reviewing on Metacritic and on lexical diversity, anti-bombers behaved similarly to bombers and that in some cases they admitted to having biased their own rating score to counter-balance the effect of RB. Considering that TLOU2 is the most reviewed item on Metacritic and the largest case of RB, it is possible that many of these anti-bombers would have not reviewed TLOU2 if not after noticing the RB.

The aim of this second analysis is to determine the prevalence of recurrent themes in the English corpus. A \lq bag-of-word' technique is employed \citep{gentzkow_text_2019}. The labelling involves the following steps: (i) theoretical identification of labels of interest; (ii) for each label, a vocabulary of tokens is constructed; and (iii) labels are assigned to reviews containing at least one token from the vocabulary o the label. For each review, a label is assigned or it is not, hence labels work as dummy variables to signal that the review mentioned the theme identified by the label.

\subsection{Identification of the Labels}

Two key themes emerged in the debate surrounding TLOU2. The first pertains to the politicisation of the game, and specifically through an over-representation of LGBTQ narrative. The second theme accounts for the opinion that professional video game critics are biased and that the industry behaves unethically.

The first theme has been unpacked in two labels: one centred around political jargon, and another centred on slurs and derogatory terminology for LGBTQ people.

The second theme is of a more complex nature and revolves around the perceived bias and dishonesty of journalists and experts. This theme can be further broken down into a broader and more abstract category, which is identified as `meta-opinion'. It refers to an opinion that is expressed to voice one's concern that the claim from someone else is deceitful and not sincere. The original concern of \#GamerGate is a `meta-opinion' because it is a protest not strictly aimed at the quality of products of the video game industry, but at the perceived lack of moral integrity of mainstream, professionalised journalists. Those who truly believe in the \#GamerGate would believe that journalists are a source of disinformation about video games. However, in Materials has been established that anti-bombers believe that it doesn't matter if the experts are wrong, but that the bombers are those spreading misinformation. Anti-bombers are also expressing a `meta-opinion' as an act of reactive communication because they warn that bombers are trying to persuade the public opinion to believe in wrong or unsubstantiated facts. The ambiguity here is that, in order to do so, they end up not looking too different from the object of their accusation.

In addition to the previous three labels, a fourth label is proposed to recognise when a textual review mentions Technical details of the video game. The vocabulary for this label includes tokens related to the quality of graphics, music, voice acting, storytelling, and character names. This label serves as a counterfactual: when it manifests alone, it signals that the user expressed solely an interest in technical details and not in the controversy surrounding the game.

\subsection{Vocabulary construction}

Four labels are identified: jargon of Politics ($P$), LGBTQ ($Q$), Meta-opinions ($M$), and Technical jargon ($T$). For each label, a \textit{prior} vocabulary is identified by selecting words that are commonly associated with that jargon. Vocabularies are then algorithmically expanded using the following method:

\begin{enumerate}
    \item The English corpus is pre-processed: all characters are low-cased and stop-words are removed.
    \item Reviews that contain any tokens that are already present in the vocabulary are removed, and only the filtered sample is kept for analysis.
    \item The list of the 2000 most frequent tokens in the filtered sample is examined. Any previously unidentified tokens that are semantically related to the vocabulary are added as a \textit{posterior} component. Tokens may be added to the vocabulary in a stemmed form to include variants of the same token.
    \item .2 and .3 are reiterated with the expanded vocabulary until none of the 2000 most frequent tokens in the filtered sample is semantically related to the label.
\end{enumerate}

The procedure is iterated for the four vocabularies (see Tables \ref{DictioP}, \ref{DictioQ}, \ref{DictioM}, and \ref{DictioH} in the Appendix). This algorithmic procedure, although expensive in terms of specification of the \textit{posterior} component of the vocabularies, allows great control and transparency.

\subsection{Results}

In Table \ref{tab:topics} are reported statistics on the four labels.

\begin{table}[h!]
\caption{Scoring behaviour across labels}
\label{tab:topics} 
\centering
\begin{tabular}{lrrrrr}
\hline\noalign{\smallskip}
Label & n & $\bar{x}$ & $f(x = 10)$ & $f(x < 2)$ \\ 
\noalign{\smallskip}\hline\noalign{\smallskip}
P & 7126 & 3.05 & 0.15 & 0.55 \\ 
Q & 5039 & 4.20 & 0.24 & 0.43 \\ 
M & 20060 & 5.93 & 0.39 & 0.27 \\ 
T & 36547 & 5.06 & 0.30 & 0.34 \\ 
\noalign{\smallskip}\hline\noalign{\smallskip}
\end{tabular}
\end{table}

Some initial observations are evident from the data: users who discussed Politics in their reviews of TLOU2 brought the most negative sentiments towards the video game. This finding is consistent with the expectation that the majority of users who reviewed the game as a political statement were politically conservative. There is a predominance of negative sentiments towards LGBTQ themes, but this label is not very frequent. Meta-opinions were much more widespread. Among this group, the average rating score was close to $6$. This is the only user group where the frequency of $x=10$ (corresponding to very positive sentiment in Fig. \ref{fig:time}) outweighed the frequency of $x<2$ (very bad sentiment), too.

Given the structure of rating scores, it is relatively easy to distinguish explicit review bombers from anti-bombers. For example, the expression of value-based judgement (ideological or `meta') paired with a $x<2$ or a $x=10$ qualifies a user as a bomber or as someone who has been influenced by RB. Yet, users with value-based motivations bombers could still conceal their true intentions\footnote{A deceptive behaviour coherent with customs associated to 4chan's users \citep{knuttila_user_2011,ludemann_digital_2021}.} mentioning technical aspects of the video game only, to appear more valid or authoritative and to mask their real concerns about the political implications of the User Score. Indeed, Over half of the reviews in the corpus mentioned Technical jargon, but almost two-thirds of these ``technical" reviews still adopted extreme scores ($0.3 + 0.34 = 0.64$). An alternative explanation is that users cared about a good technical product but were disappointed that it did not align with their political values.

\subsection{Four clusters of users}

The English Corpus of reviews can be partitioned partition in four clusters, according to a criterion of `rhetorical objectivity':

\begin{enumerate}
    \item The less rhetorically objective cluster consists in those reviews where all 4 themes are absent.
    \item Next are the reviews mentioning one of the themes P, Q, or M, but not T.
    \item These are followed by reviews mentioning one across P, Q, or M, plus T is present.
    \item The most rhetorically objective cluster contains the reviews where exclusively T and other themes are not present.
\end{enumerate}

\begin{figure}[hbt!]
	\centering
		\includegraphics[width=12cm]{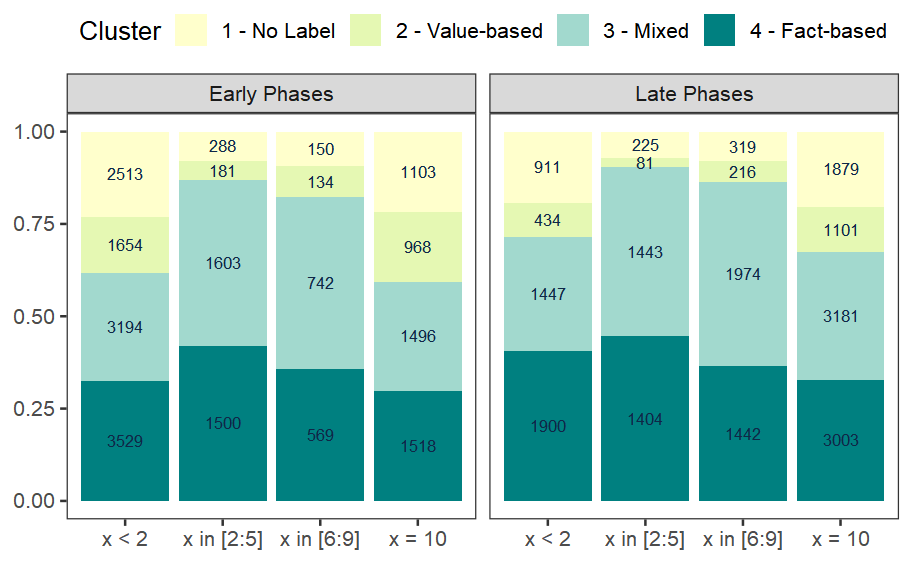}
	\caption{\textbf{Prevalence of clusters.} \\ The height of the stacked bars represents the prevalence of the cluster among that class of scoring behaviour. The number within it is the absolute number, instead.}
	\label{fig:bp}
\end{figure}

These clusters are ordered only by a `rhetorical' objectivity because it is assumed that users may avoid mentioning non-Technical themes to sound more objective. In Fig. \ref{fig:bp} the prevalence of clusters is crossed with scoring behaviour, in the first days (early two phases) and in the late days (late two phases).

In terms of absolute values, as expected (Fig. \ref{fig:time}), the scoring behaviour shifts towards higher scores, but in terms of relative values there are more differences across rating behavior than over time. In the late stages, the prevalence of the Fact-based cluster increased. The most effective negative reviews ($x < 2$) were those that shifted more into fact-based expressions of judgement. An explanation is that, as their absolute numbers went down, bombers acknowledged that a communication campaign built on political conservativism and anti-LGTBQ was backfiring and that anti-bombers were countering the effectiveness of the RB in pushing low the User Score. Maybe bombers adopted a new narrative based on the technical poverty of the video game. Pure value-based opinions and low-informative reviews (No Label) were rather scarce in non-extreme rating scores, although purely fact-based reviews did not increase in late phases among these.

\begin{figure}[hbt!]
	\centering
		\includegraphics[width=12cm]{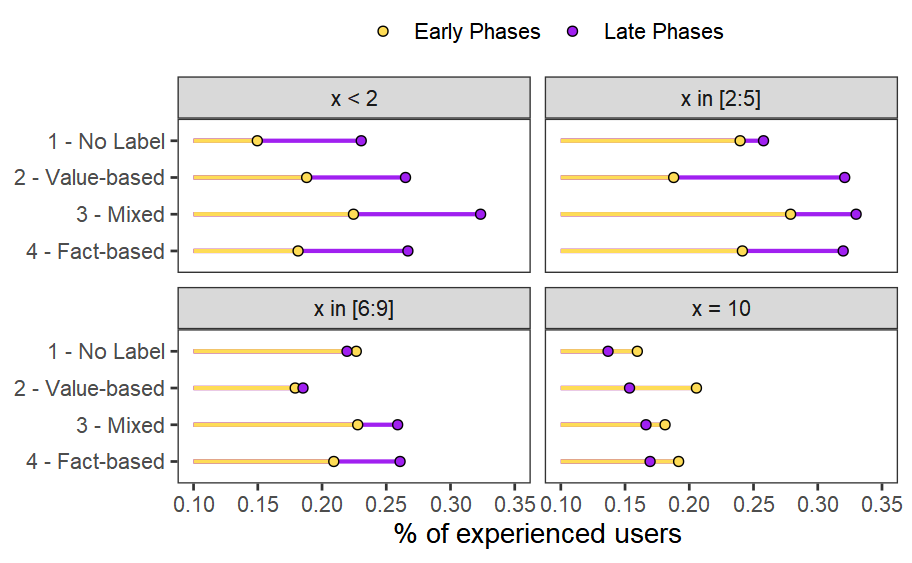}
	\caption{\textbf{Differences in the share of expert users across clusters.} The purple line measures how $K=1$ increased in the last two phases, compared to the first two. When the purple dot is on the left of the yellow one, it decreased.}
	\label{fig:fk}
\end{figure}

The rise of anti-bombers should not be left unchecked in the cluster analysis. Indeed, in Fig \ref{fig:fk}, $x = 10$ is the only scoring behaviour that over time, reduced the prevalence of accounts with a previous rating experience in Metacritic ($K$). This fact is concerning because this result is coherent with a genuine interest of more people in Metacritic, but it could also signal the employment of \textit{sock-puppets} or even a post-RB campaign of \textit{astroturfing}.

The base rate and the positive shift in the share of experienced users with a negative sentiment across all clusters in Fig. \ref{fig:fk} confirms that the RB was not an isolated ploy from a small group of 4chan `trolls'. It was more a `call to arms' that resonated with a cluster of Metacritic users'. This is unlikely an effect of the overall reduction of negative reviewers since the absolute number of users who rated TLOU2 between $2$ and $5$ (very low scores for Metacritic) did not meaningfully reduce in the late phases.

\begin{figure}[hbt!]
	\centering
		\includegraphics[width=12cm]{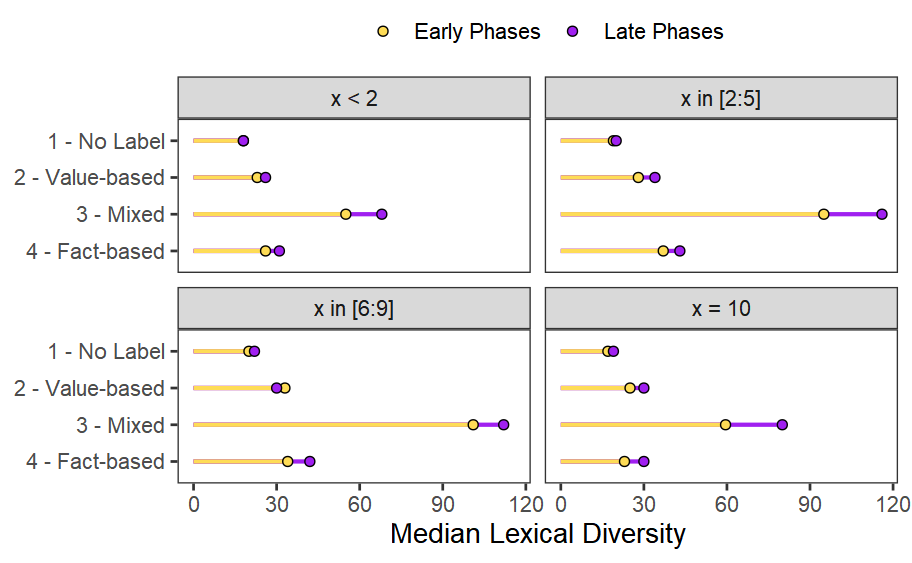}
	\caption{\textbf{Differences in median lexical diversity across clusters.} The purple line measures how the median $D$ increased in the last two phases, compared to the first two. When the purple dot is on the left of the yellow one, it decreased.}
	\label{fig:ld}
\end{figure}

An extensive statistical summary across all the combinations of labels and an analogous representation of Fig. \ref{fig:ld} within clusters 2 and 3 only are presented in Appendix C. These supplementary results confirm preliminary results in Table \ref{tab:topics}: bombing behaviour is associated with ideological expressions, anti-bombing behaviour is associated with meta-opinions, and proportions of reviews across clusters are rather stable over time. An interpretation of this stability is that past reviews did not really affect future the opinion of future reviewers, it only affected the probability to bring anti-bombers, possibly previously uninterested in Metacritic, into reviewing positively TLOU2 to contrast RB. If these results demonstrate that a cluster of ideologically-driven users pre-existed among Metacritic's user demography, a mirrored argument can be proposed for anti-bombers: these are not people with an interest in Metacritic, but are people who felt morally or ideologically obliged to contrast RB, with no evidence of massive adoption of \textit{sock-puppets}.

\section{Discussion}

The RB of TLOU2 reflects clearly the existence of groups of people that, once nudged by the buzz of the media, adopted an open platform of online review to express their ideology-driven opinions. A very coarse reconstruction of what happened is that many bombers held conspiratorial-looking (although, consistent within their worldview) beliefs that progressive forces were acting to spoil a shared identity. Immediately after, people who were not interested in Metacritic (or, even in the video game), felt the moral imperative to fight back against what they perceived as an unfair and possibly homophobic attempt to push a politically reactionary message through the platform Metacritic.

According to analysis, the evidential impact of fake reviews is substantially small. Instead, these two factions are representative of two ideological cultures which, to simplify it, descend from the conventional dichotomy of the political spectrum between moral conservativism and moral progressivism. Both factions represent a specific idea of morality, and both factions felt morally legitimised to (de)bias the User Score for a higher ethical end.

As shown by Fig. \ref{fig:bomb}, after the `TLOU2 incident', review bombs intensified in frequency over time. It means that between 2020 and 2022 users of online reviews platforms learnt that their message can be effectively conveyed in media through RB. While the evidence is relatively scarce on the topic, findings confirm that RB is driven by an ideological, ethical, or moral grudge against the item or against other people. In other words, if RB will be a future trend, there will be spikes in value-based reviews. The premise for the following recommendations is that at the current stage platforms of online reviews are ill-equipped to deal with an incoming surge of value-based reviews.

An immediate solution to RB would be the aforementioned suppression of value-based reviews in summary statistics of the item, and in the scores for rankings. This solution is rooted in the belief that an objective quantification of a latent trait concerns the summarisation of multiple observations of facts, and moral judgements have nothing to do with online reviews. While this strategy is theoretically consistent, in practice the solution of suppressing value-based brings to a methodological issue: in Fig. \ref{fig:ld} is shown that Mixed (Technical and Ideological) reviews are among the most lexically rich while being not infrequent at all (see Fig. \ref{fig:bp}, and Fig. \ref{fig:ideo} in the Appendix D). In other words, if reviews based only on moral merit are akin to `spam reviews', once the moral judgement is combined with a recognition of the technical merits of the item, users will very likely spend a serious effort to justify their opinion, with only a minor (quadratic) dependence to the score. Should then only reviews in clusters 1 and 2 be suppressed or shrunk in summary statistics as the User Score? It is still an inconvenient solution because, as shown in the section Thematic Labelling, astute bombers can find their way to  circumvent such a countermeasure.

A first proposal for managing a surge of value-based reviews as an effective method to voice moral distress is inspired by methodological issues in the detection of \textit{sock-puppets} among users who never reviewed before ($K = 0$). The recommendation is to allow users to rate an item before its release, and then allow them to re-rate it after the release, too, preserving their previous rating. A rating before the release quantifies an expectation towards the future outcome. The immediate feature that is introduced is that \textit{pre} and \textit{post} scores can be compared within the same user, so it can be quantified how high the expectations are for a future item, and when these are met or not. There is another perk of collecting pre-release ratings: it makes it harder to organise a Review Bomb through \textit{sock-puppets}: after a release of a relevant item such as TLOU2 there is a peak in user registrations which can be due to the media exposure or for other reasons; a peak in registration \textit{before} a release would be much more suspicious instead, and facilitate detection of coordinated \textit{sock-puppets}.

The second proposal originates from the recognition of meta-opinions across the value-based and mixed reviews. They testify how, under the correct condition, there is an untapped potential in people's desire to judge and evaluate how others rate items. A simple implementation would be to allow users to actually attach their answers to a pre-existing review. This is not recommended, because it would be dispersive: if one has a general meta-opinion (e.g. on a group), why should it be attached to a specific review? The recommendation is to allow users with a certain level of experience and fidelity to the platform to access an additional function, working as a tribunal system, instead. In a tribunal system, a user acts as a juror: a review of an item from someone else is subject to his judgement. The second-order rating of the juror contributes to a weighing criterion. In other words, a weighting criterion would emerge organically from a crowd-sourcing mechanism and not from a top-down approach designed to for disciplining unruly users.

There are two structural differences between a judgement in a tribunal system and the submission of a meta-opinion as a review. The first is that the juror does not have control over what to judge, and the second is that potentially the same juror can make multiple judgements on reviews of the same object. In this sense, an expert user can have a greater influence than a new account, but in a legitimate way, because he cannot abuse it by choosing to judge only reviews on the same object. There could be some issues, for example, that users are led to express meta-opinions only for certain items, but the problem does not really arise, because once the juror request to judge reviews for an item, then it is up to the platform to decide the criterion for assigning which review to the juror, or the number of reviews that juror can judge.

\section{Conclusions}

A limitation of this study is that, based only on explicit textual features, the analysis likely did not spot all the occurrences of fake reviews, possibly because these have been quickly suppressed by Metacritic. Connected to the issue of fake detection is the granularity of the time variable: reviews are indexed by the day, but since many reviews appeared in the first four days (Phases I and II), dynamics within the day are lost. Research on Review Bombs should account for novel techniques of fabrication through Large Language Models. These technologies can automate the production of highly sophisticated fake reviews which are not even detectable by the best detection algorithms \citep{adelani_generating_2020,stiff_detecting_2022}. A possible remedy for applications intensively focused on the quantification of fake reviews should then account for `triangulation' of different approaches to detection.

However, the main limitation of the study is that it does not involve any \textit{in vivo} counterfactual controlled intervention, so possibly all the identifies general structures (e.g. anti-bombers, meta-opinions) have only an idiosyncratic validity, limited within the observed data-generating process. The external validity of these findings should not undermine the goodness of the recommendations, yet this study cannot claim, for example, the thesis that every form of Review Bombing must necessarily be followed by an anti-bombing movement. The data-generating process does not follow a controlled experimental design, so persists an unanswered question about under what conditions a RB is followed by a wave of anti-bombers. This is a valid direction for future research and \textit{in vivo} experimental studies can attempt to demonstrate some causal determinants of the occurrence of polarisation in opinion dynamics\citep{krueger_conformity_2017,bail_exposure_2018,kvam_rational_2022}. 

The distinction between fact-based and value-based opinions can be related to two traditional paradigms of the theory of evaluation \citep{stufflebeam_evaluation_2007}:

\begin{itemize}
    \item objectivism, which is aimed at inferring latent traits of pre-identified concepts and in this case, pre-identified \textit{items};
    \item participatory evaluation, or the idea that the data-generating process under analysis should reflect a pre-existing disposition (empowerment) of the respondent.    
\end{itemize}

Conventionally, these two paradigms, which are sometimes fully (and erroneously) conflated to quantitative empiricism and qualitative (or \textit{critical}) empiricism are seen as opposed \citep{greene_defining_1997,mcgregor_paradigm_2010}. This study highlights how this particular cleavage is reflected in online reviews: for platforms, it is important to establish that their scores are unbiased and transparent, as predicated by an objective evaluation. Yet, scores are aggregations of opinions from a crowd of participants who feel empowered when they rate items. For those users, it is not a paid job, hence users leave honest and authentic reviews for reasons other than financial gain, and possibly they feel an emotional connection with their role of opinion providers ("My opinion counts!").

Although the platform aims to present itself as objective and independent, in practice its business model is based on tapping into a pre-existing predisposition towards participation in evaluative processes \citep{napoli_audience_2012, livingstone_participation_2013, zijlstra_folk_2019}. The study shows that Review Bombing is often empowering for the user, even if it conflicts with the platform's goals, and that this empowerment can undermine the claim that platform scores are objective. The conclusion is that the problem posed by the Review Bomb is not limited to an issue of disinformation. The fundamental problem is of misalignment between the interests of the platform and those of a relevant portion of its users.

Research on online reviews, despite being two-decade lasting as a specific framework of information management, is not integrated with the theoretical debate on evaluation methodologies. The emergence of Review Bomb should therefore be seen as an opportunity for further theoretical developments in this field of information management\citep{schwandt_judging_2007,venkatesh_bridging_2013,das_value_2020}. In this regard, while commonly found in social media, the concept of meta-opinion is currently foreign to research on online reviews because it goes against the standard model of users leaving reviews on items. Indeed, meta-opinions still express a point of view that is not strictly about the item itself but connects previous opinions about the item, so their function in the communication is to provide arguments to refute or corroborate, hence to weight, the trustworthiness of a previous message.

Finally, the distinction between fact-based and value-based opinions, and the concept of meta-opinions have the potential to enrich theories in various fields, such as Natural Language Processing (NLP), Psychology, Communication, and Political science. For example, the detection of the ideological charge behind a short text may be a step toward a better understanding of the hidden motivations behind it. The scoring system of online reviews is helpful because the scoring behaviour is often strategical, making it easier to infer a motivation.


\newpage

\appendix

\section{Detailed Summary}

In Table \ref{tab:scores} are reported the frequencies of the scores ($x$) in the English corpus, along with the median lexical diversity ($D$) in the class of score, and the relative frequency of users who have already submitted a review on Metacritic before TLOU2.

\begin{table}[hbt!]
\caption{Scores, in absolute and relative frequency, their median lexical diversity ($D$) and the frequency of users with previous reviews}
\label{tab:scores}
\centering
\begin{tabular}{lrrrr}
\noalign{\smallskip}\hline\noalign{\smallskip}
$x$ & $n(x)$ & $f(x)$ & $Med(D)$ & $f(k_u > 0)$ \\
\noalign{\smallskip}\hline\noalign{\smallskip}
0 & 13969 & 0.27 & 27 & 0.20 \\ 
  1 & 4923 & 0.10 & 33 & 0.23 \\ 
  2 & 2769 & 0.05 & 48 & 0.25 \\ 
  3 & 2391 & 0.05 & 57 & 0.27 \\ 
  4 & 1950 & 0.04 & 71 & 0.30 \\ 
  5 & 1234 & 0.02 & 68 & 0.31 \\ 
  6 & 885 & 0.02 & 87 & 0.33 \\ 
  7 & 822 & 0.02 & 70 & 0.34 \\ 
  8 & 1660 & 0.03 & 64 & 0.22 \\ 
  9 & 3399 & 0.07 & 53 & 0.18 \\ 
  10 & 17118 & 0.33 & 33 & 0.16 \\  
\noalign{\smallskip}\hline\noalign{\smallskip}
\end{tabular}
\end{table}

\newpage

\section{Vocabularies}

\begin{table} [hbt!]
\caption{Dictionary for Politics - P}
\label{DictioP} 
\centering
\begin{tabular}{p{2cm}p{9cm}}
\hline\noalign{\smallskip}
& Tokens \\
\noalign{\smallskip}\hline\noalign{\smallskip}
Prior & agenda, alt right, altright, cancel cult, cancell, conservative, democra, far left, far right, fascis, feminis, gamergate, ideol, jew, kike, leftis, nazi, politic, progressive, propagand, racis, shill, sjw, social justice warrior, virtue sign\\
\noalign{\smallskip}\hline\noalign{\smallskip}
Posterior & activis, alt-right, anita, asia, far-right, feminaz, freedom of, globohomo,  idealog, idelog, ideolog, lectur, moral, polical, propogan, religio, retcon, socialis, sponsor, trump, white man, white men, woke\\
\noalign{\smallskip}\hline\noalign{\smallskip}
\end{tabular}
\end{table}

\begin{table} [hbt!]
\caption{Vocabulary for LGBTQ - Q}
\label{DictioQ} 
\centering
\begin{tabular}{p{2cm}p{9cm}}
\hline\noalign{\smallskip}
& Tokens \\
\noalign{\smallskip}\hline\noalign{\smallskip}
Prior & gender, bisex, dyke, fag, faggot, gay, gender, heterosex, homophob, homosex, intersex, lesb, lgbt, non-binary, nonbinary, pansexual, queer, trann \\
\noalign{\smallskip}\hline\noalign{\smallskip}
Posterior & androge, cis, degenerate, dyke, erotic, femenin, hetero, homos, hulk, inclusi, kiss, lbgt, lezb, lezz, masculin, pedo, porn, same sex, sex scene, shemale, sodom, stereotyp, taboo \\
\noalign{\smallskip}\hline\noalign{\smallskip}
\end{tabular}
\end{table}

\begin{table} [hbt!]
\caption{Vocabulary for Meta-opinions - M}
\label{DictioM} 
\centering
\begin{tabular}{p{2cm}p{9cm}}
\hline\noalign{\smallskip}
& Tokens \\
\noalign{\smallskip}\hline\noalign{\smallskip}
Prior & bombin, boycot, controvers, critic, fake, journalis, metacritic, ratin, sabotag, scor, streamer, troll \\
\noalign{\smallskip}\hline\noalign{\smallskip}
Posterior & 19th, are mad, balanc, bandwag, bias, blind, bots, bottin, brigad, comment, communit, complain, critiq, crybab, divisiv, downvot, fanboy, first day, grade, hater, ignore the, immature, incel, jedi, moron, overreact, people who, polar, salty, statistic, star war, the 0, the zero, user, who hate \\
\noalign{\smallskip}\hline\noalign{\smallskip}
\end{tabular}
\end{table}

\begin{table} [hbt!]
\caption{Vocabulary of Technical jargon - T}
\label{DictioH} 
\centering
\begin{tabular}{p{2cm}p{9cm}}
\hline\noalign{\smallskip}
& Tokens \\
\noalign{\smallskip}\hline\noalign{\smallskip}
Prior & abbie, actin, actor, antagonist, boss, character, dinah, ellie, fireflies, gameplay, gold, graphic, hero, jess, jj, joel, lev, manni, mechanic, murderer, music, narrat, protagonist, storytell, tomm, villain, visua, writing, yara \\
\noalign{\smallskip}\hline\noalign{\smallskip}
Posterior & animat, atmospher, bugs, cinematic, clich, collectibl, combat, cut scene, cutscen, design, dialog, ebby, environment, execut, flashbac, flaw, frame rat, framerat, game play, gamebreak, gaming exp, gampl, glitc, gore, goty, grafic, improvemen, innovative, linear, loot, melee, motion blur, open world, openworl, pathin, performa, platin, plot, puzzle, realistic, sandbox, script, sideque, storyl, structur, technic, worldbuild \\
\noalign{\smallskip}\hline\noalign{\smallskip}
\end{tabular}
\end{table}

\newpage

In certain cases, bigrams have been included to enhance the meaning of individual tokens. For example, the token ``0" was a frequent occurrence, but it did not make much sense on its own. However, when combined with the definite article ``the" to form the bigram ``the 0", it becomes a clear signal that people are referring to those users who have assigned a score of 0 to the game.

\subsection{Weird tokens}
To illustrate the logic behind the iterative process of building up vocabularies, some unusual tokens are discussed. For example, the words ``jedi" and ``star war" were mentioned, respectively, in 305 and 298 reviews in the English corpus. These terms specifically refer to the movie ``Star Wars: The Last Jedi": many users drew similarities not between the movie and the video game, but between the debates about them. In both cases, some critics denounced an attempt to force a feminist narrative into a fictional genre. Thus, these tokens were included as examples of meta-opinions.

Instead, ``anita" (256 reviews) refers to Anita Sarkeesian, a video game critic. This token was associated to political criticism because Sarkeesian has been a being a target of the \#GamerGate campaign \citep{braithwaite_its_2016}.

Some tokens are misspelled versions (e.g., ``prop\textit{o}ganda" for ``prop\textit{a}ganda"), while others were strongly rooted in niches of political jargon: a ``npc" (206 reviews) is someone who follows mainstream uncritically; ``sjw" (1345 reviews) is an acronym for ``social justice warrior"; ``hulk" is a derogatory term for female bodybuilders, in this context it made sense to associate it with slurs against ``queer" people because female bodybuilders do not conform to traditional gender aesthetics.

There is one recurrent word with am ambiguous semantics, which is ``xbox" (236 reviews). It refers to a competitor product and it was mentioned in negative reviews suggesting to shift to competitors. This theme has not been regarded as related to any of the four labels. All similar cases of ambiguity have a lesser frequency in the English corpus.

\newpage

\section{Disaggregated clusters}

Tables \ref{tab:dept0} and \ref{tab:dept1} provide an in-depth statistical summary across the four thematic labels. Each combination of them has been disaggregated into separate rows in the table. In Table \ref{tab:dept0} the $x$ score of the review is re-coded as a categorical variable: when $x < 6$, Sentiment is Negative, when $x > 6$, it is Positive. In Tables \ref{tab:dept1}, the statistics are computed for reviews in Phases 1 and 2 (Early phases, the first four days), and second, accounting only for reviews in Phases 4 and 5 (Late phases).

\begin{table}[hbt!]
\centering
\caption{Sentiment and effort across thematic labels}
\label{tab:dept0}
\begin{tabular}{lllllrrr}
\hline\noalign{\smallskip}
P & Q & M & T & Sentiment & n & $Med(D)$ & $f(K=1)$ \\ 
\noalign{\smallskip}\hline\noalign{\smallskip}
0 & 0 & 0 & 0 & Bad & 4736 & 18 & 0.18 \\ 
  0 & 0 & 0 & 0 & Good & 4089 & 19 & 0.15 \\ 
  1 & 0 & 0 & 0 & Bad & 787 & 21 & 0.19 \\ 
  1 & 0 & 0 & 0 & Good & 81 & 30 & 0.15 \\ 
  0 & 1 & 0 & 0 & Bad & 243 & 19 & 0.14 \\ 
  0 & 1 & 0 & 0 & Good & 66 & 25 & 0.12 \\ 
  0 & 0 & 1 & 0 & Bad & 1072 & 26 & 0.21 \\ 
  0 & 0 & 1 & 0 & Good & 2431 & 27 & 0.17 \\ 
  0 & 0 & 0 & 1 & Bad & 10219 & 31 & 0.22 \\ 
  0 & 0 & 0 & 1 & Good & 7482 & 30 & 0.18 \\ 
  1 & 1 & 0 & 0 & Bad & 102 & 24 & 0.22 \\ 
  1 & 1 & 0 & 0 & Good & 13 & 25 & 0.15 \\ 
  1 & 0 & 1 & 0 & Bad & 394 & 30 & 0.23 \\ 
  1 & 0 & 1 & 0 & Good & 118 & 41 & 0.21 \\ 
  1 & 0 & 0 & 1 & Bad & 1388 & 40 & 0.28 \\ 
  1 & 0 & 0 & 1 & Good & 205 & 58 & 0.20 \\ 
  0 & 1 & 1 & 0 & Bad & 95 & 30 & 0.17 \\ 
  0 & 1 & 1 & 0 & Good & 139 & 35 & 0.17 \\ 
  0 & 1 & 0 & 1 & Bad & 683 & 47 & 0.19 \\ 
  0 & 1 & 0 & 1 & Good & 189 & 45 & 0.17 \\ 
  0 & 0 & 1 & 1 & Bad & 4066 & 75 & 0.26 \\ 
  0 & 0 & 1 & 1 & Good & 6186 & 76 & 0.19 \\ 
  1 & 1 & 1 & 0 & Bad & 123 & 31 & 0.17 \\ 
  1 & 1 & 1 & 0 & Good & 32 & 45 & 0.28 \\ 
  1 & 1 & 0 & 1 & Bad & 314 & 63 & 0.24 \\ 
  1 & 1 & 0 & 1 & Good & 46 & 67 & 0.22 \\ 
  1 & 0 & 1 & 1 & Bad & 1387 & 91 & 0.30 \\ 
  1 & 0 & 1 & 1 & Good & 665 & 130 & 0.20 \\ 
  0 & 1 & 1 & 1 & Bad & 762 & 114 & 0.25 \\ 
  0 & 1 & 1 & 1 & Good & 786 & 102 & 0.19 \\ 
  1 & 1 & 1 & 1 & Bad & 865 & 150 & 0.27 \\ 
  1 & 1 & 1 & 1 & Good & 471 & 188 & 0.19 \\
\noalign{\smallskip}\hline\noalign{\smallskip}
\end{tabular} 
\end{table}

\begin{table}[h!]
\centering
\caption{In-depth statistics on effort across thematic labels}
\label{tab:dept1}
\begin{tabular}{lllllrrrr}
\hline\noalign{\smallskip}
P & Q & M & T & Phases & $n$ & $\bar{x}$ & $Med(D)$ & $f(K=1)$  \\ 
\noalign{\smallskip}\hline\noalign{\smallskip}
0 & 0 & 0 & 0 & Early & 4054 & 3.36 & 18 & 0.16 \\ 
  0 & 0 & 0 & 0 & Late & 3334 & 6.72 & 19 & 0.18 \\ 
  1 & 0 & 0 & 0 & Early & 559 & 0.96 & 21 & 0.17 \\ 
  1 & 0 & 0 & 0 & Late & 189 & 2.28 & 22 & 0.28 \\ 
  0 & 1 & 0 & 0 & Early & 222 & 2.04 & 19 & 0.14 \\ 
  0 & 1 & 0 & 0 & Late & 36 & 6.00 & 22 & 0.14 \\ 
  0 & 0 & 1 & 0 & Early & 1496 & 5.98 & 26 & 0.21 \\ 
  0 & 0 & 1 & 0 & Late & 1398 & 8.22 & 29 & 0.17 \\ 
  0 & 0 & 0 & 1 & Early & 7116 & 3.54 & 28 & 0.20 \\ 
  0 & 0 & 0 & 1 & Late & 7749 & 6.05 & 34 & 0.24 \\ 
  1 & 1 & 0 & 0 & Early & 82 & 1.45 & 25 & 0.16 \\ 
  1 & 1 & 0 & 0 & Late & 14 & 2.93 & 21 & 0.36 \\ 
  1 & 0 & 1 & 0 & Early & 326 & 1.94 & 30 & 0.21 \\ 
  1 & 0 & 1 & 0 & Late & 115 & 4.09 & 38 & 0.28 \\ 
  1 & 0 & 0 & 1 & Early & 876 & 1.37 & 39 & 0.24 \\ 
  1 & 0 & 0 & 1 & Late & 473 & 3.39 & 47 & 0.35 \\ 
  0 & 1 & 1 & 0 & Early & 142 & 5.42 & 30 & 0.20 \\ 
  0 & 1 & 1 & 0 & Late & 60 & 7.57 & 34 & 0.17 \\ 
  0 & 1 & 0 & 1 & Early & 479 & 2.33 & 44 & 0.17 \\ 
  0 & 1 & 0 & 1 & Late & 257 & 4.21 & 57 & 0.25 \\ 
  0 & 0 & 1 & 1 & Early & 3338 & 5.01 & 65 & 0.23 \\ 
  0 & 0 & 1 & 1 & Late & 5211 & 7.36 & 83 & 0.23 \\ 
  1 & 1 & 1 & 0 & Early & 110 & 1.91 & 31 & 0.20 \\ 
  1 & 1 & 1 & 0 & Late & 20 & 5.60 & 53 & 0.20 \\ 
  1 & 1 & 0 & 1 & Early & 211 & 1.44 & 59 & 0.22 \\ 
  1 & 1 & 0 & 1 & Late & 95 & 3.72 & 81 & 0.27 \\ 
  1 & 0 & 1 & 1 & Early & 928 & 2.72 & 87 & 0.26 \\ 
  1 & 0 & 1 & 1 & Late & 827 & 5.51 & 128 & 0.28 \\ 
  0 & 1 & 1 & 1 & Early & 619 & 4.73 & 97 & 0.21 \\ 
  0 & 1 & 1 & 1 & Late & 663 & 6.62 & 123 & 0.23 \\ 
  1 & 1 & 1 & 1 & Early & 584 & 3.40 & 134 & 0.21 \\ 
  1 & 1 & 1 & 1 & Late & 519 & 5.61 & 191 & 0.29 \\ 
\noalign{\smallskip}\hline\noalign{\smallskip}
\end{tabular} 
\end{table}

It is evident in Table \ref{tab:dept1} that the proposed clustering scheme is representative of the most frequent disaggregate clusters. For example, the disaggregated cases when $T = 1$ and all the other labels are absent are the most frequent in the corpus.

Through the disaggregated clusters it is possible to notice that within the Mixed cluster is relevant the case when $T=1$, $M=1$ and the other two labels are absent. This is the case that registered the highest rise in the number of reviews $n$ between Early and Late phases: from $3338$ to $5211$, with a conspicuous rise of $\bar{x}$ and no shifts in $f(k_u = 1)$. This rise can be easily seen as a shred of consistent evidence for the `counter-RB' effect of anti-bombers. It can be contrasted with the case of $M = 1$ and no co-occurrences: in this case, to an impressive rise of positive sentiment in Late phases, $f(k_u = 1)$ dropped significantly.

\begin{figure}[hbt!]
	\centering
		\includegraphics[width=12cm]{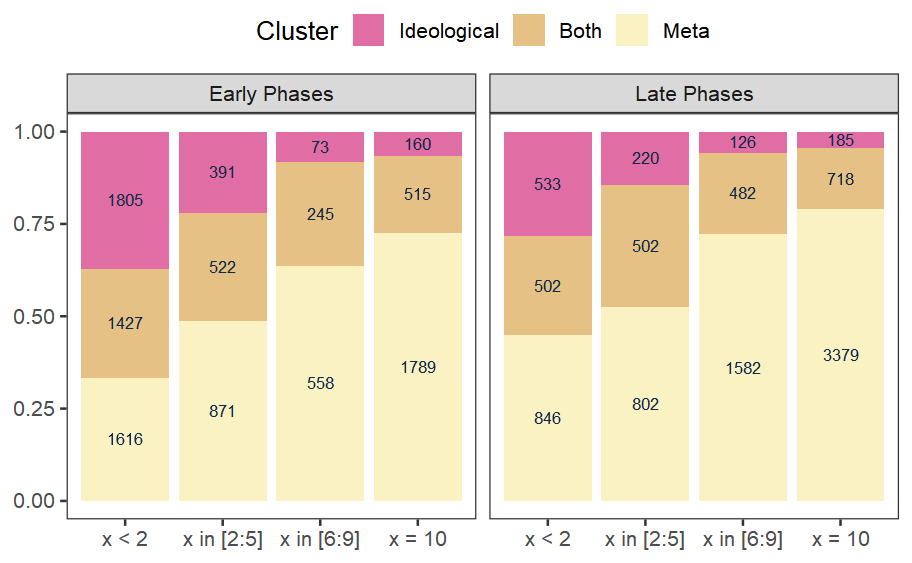}
	\caption{\textbf{Classes of reviews within the non-Technical labels.}\\ In dark pink, the reviews mentioning only Political or LGTBQ keywords, in pale yellow those mentioning only Meta-opinions but without an ideological charge.}
	\label{fig:ideo}
\end{figure}

Fig. \ref{fig:ideo} represents the shares of sub-classes of reviews within all the reviews mentioning at least one of the 3 value-based labels (Politics, LGTBQ and Meta-opinions), or value-based reviews. The difference between the first four days (Early Phases) and Late phases is not striking. It is worth mentioning, instead, that scores between $6$ and $9$, a frequent scoring behaviour in Metacritic, behaved more similarly to extreme positive scoring ($x = 10$) than those who scored between $2$ and $5$ to extreme negative scoring ($x < 2$). The conclusion of this result is that RB has been considered a factor worth being mentioned even among those users who were less likely to bias their own score as a (counter)effect of RB.

\newpage

\section{Analysis of fakes}

In Table \ref{tab:fake} are compared the 1722 detected fake reviews with the others in the English corpus. These reviews mostly differ in fakes being prevalent in the early phases (first 4 days), and slightly more negative.

\begin{table}[hbt!]
\caption{Statistics on detected fakes.}
\label{tab:fake} 
\centering
\begin{tabular}{lrrrrrr}
\hline\noalign{\smallskip}
Fake & $n$ & $Med(D)$ & $f(K=1)$ & $\bar{x}$ & Early & Late \\ 
\noalign{\smallskip}\hline\noalign{\smallskip}
Yes & 1722 & 37 & 0.21 & 4.02 & 0.45 & 0.36 \\
No & 49398 & 36 & 0.21 & 5.08 & 0.41 & 0.41 \\ 
\noalign{\smallskip}\hline\noalign{\smallskip}
\end{tabular}
\end{table}

The characterisation of fakes is possible more coherent with just `spam reviews' or just `nonsense', since these are mostly filled with nonsensical messages or usernames. Anyway, results should not be considered a conclusive inference on the latent behaviour of people who employed \textit{sock-puppets} throughout the Review Bomb. Both an attentive human imputer and a modern automated system (e.g. based on GPT technology) can make fake reviews that are it difficult to spot, as well as grammatically flawless and with unique stylistic differentiation while spreading fake content. Finally, the impact of false positives should be accounted for in the interpretation of Table \ref{tab:fake}.

\end{document}